# Molding Wetting by Laser-Induced Nanostructures

Aleksander G. Kovačević,[1,*] Suzana Petrović,[2] Alexandros Mimidis,[3] Emmanuel Stratakis,[3] Dejan Pantelić [1] and Branko Kolaric [1,4]

[1] Institute of Physics, University of Belgrade, Pregrevica 118, 11080 Belgrade, Serbia;

[2] Institute of Nuclear Sciences "Vinča", University of Belgrade, Mike Petrovića Alasa 12-15, 11351 Vinča-Belgrade, Serbia;

[3] Institute for Electronic Structure and Laser, FORTH, 100 Nikolaou Plastira, 10073 Heraklion, Greece;

[4] Micro- and Nanophotonic Materials Group, University of Mons, Place du Parc 20, 7000 Mons, Belgium

[*] Correspondence: Aleksander.Kovacevic@ipb.ac.rs.

**Abstract**: The influence of material characteristics— i.e., type or surface texture—to wetting properties is nowadays increased by the implementation of ultrafast lasers for nanostructuring. In this account, we exposed multilayer thin metal film samples of different materials to a femtosecond laser beam at a 1030 nm wavelength. The interaction generated high-quality laser-induced periodic surface structures (LIPSS) of spatial periods between 740 and 790 nm and with maximal average corrugation height below 100 nm. The contact angle (CA) values of the water droplets on the surface were estimated and the values between unmodified and modified samples were compared. Even though the laser interaction changed both the surface morphology and the chemical composition, the wetting properties were predominantly influenced by the small change in morphology causing the increase in the contact angle of ~80%, which could not be explained classically. The influence of both surface corrugation and chemical composition to the wetting properties has been thoroughly investigated, discussed and explained. The presented results clearly confirm that femtosecond patterning can be used to mold wetting properties.

## 1. Introduction

Generating periodical sub-wavelength structures on material surfaces by interaction with pulsed laser beams has been performed known for some time [1–4]. The structures in the form of parallel ripples—laser-induced periodic surface structures (LIPSS)—have been reported on various materials, including dielectric materials, metals, semiconductors, and graphene [5–7]. Being in the nanometer scale, and with the occurrence due to the laser-surface interaction, they are a frequent subject of investigation in nano-optics (nanophotonics).

The causes of LIPSS generation and shaping are seen most probably in the emergence of the surface plasmon polariton (SPP), as well as in the hydrodynamic features [8–10]. If compared to single layer metal films, more regular LIPSS are generated with low-fluence femtosecond (fs) laser beam interaction with multilayer thin metal films, since the existence of the metal sublayer influences the quality and stability of LIPSS [11].

Changing the wetting and tribological properties of the material by LIPSS formation opens new fields of application in nano/microfluidics, optofluidics, fluid microreactors, biomedicine, biochemical sensors, and thermal management [12,13]. The control of wetting properties and achieving super-hydrophobic surfaces by laser interaction have been reported for various materials: stainless steel, TiAl alloy, Si and materials coated with hydrophobic materials, such as chloroalkylsilane and fluoroalkylsilane [14–16]. In all mentioned cases super-hydrophobicity is achieved by the chemical modification of the surface or micro-structuring.

Laser interaction can induce the change in chemical composition of the surface, forming ultra-thin oxide layers, which contribute to the wetting. Wetting is initially enhanced, but as time passes, it is ultimately reversed by surface chemistry phenomena that take place on the irradiated surfaces—i.e., hydrophobicity is increased [16,17].

Van der Waals and electrostatic forces play an important role in adsorption, adhesion and wetting phenomena [18]. The Casimir force shows application potential in the field of micro- and nano-electromechanical systems—engineered devices, where controlling forces between microscopic bodies or surfaces are crucial for a variety of applications [19,20]. The influence of submicron surface corrugations on Lifshitz–van der Waals forces have been calculated for polyethylene, where surface nano-patterning is responsible for changing the forces from attractive to repulsive [21,22]. For metal surface nano-structuring at scales below the plasma wavelength, the Casimir interaction decreases faster than usual for large inter-surface separation, while at short separations an equivalent pressure is larger [23]. The motivation of our study is to experimentally reveal the link between wetting and nano-corrugation [21,22] and to extend the potential application of nanostructures. Surface nanostructures affect the optical characteristics via refractive index on the one hand, while on the other, they influence the wetting characteristics. Patterning the surface of the lamellar material changes the characteristics of the interface, as well as the characteristic of the bulk to some extent. The nanometer level of the patterns increases the importance of their control and places it in the field of nano-optics/nano-photonics.

The morphology of surface micro/nanostructures obtained by fs laser is evaluated with the aim of determining the influence of morphology on their wetting. By interaction with fs laser beam, we generated LIPSS on the surfaces of several metallic multilayer materials. We estimated the contact angle (CA) difference for different materials and also for materials before and after LIPSS forming, thus examining the change in hydrophobicity.

## 2. Materials and Methods

The samples were multilayer thin films on Si substrate: 15×(Zr/Ti), 15×(Ti/Zr), 8×(Zr/Cr/Ti), with the topmost layer being Zr, Ti and Zr, respectively. The total thickness of each multilayer structure was aimed to be ~500 nm, making the thicknesses of individual layers—controlled by the deposition rate—about 17 nm for bi-layer samples and ~21 nm for tri-layer sample. The samples were exposed to femtosecond beam which generated the LIPSS.

The laser source was the Pharos SP Yb:KGW laser system from Light Conversion. The surfaces of the thin films were irradiated at normal incidence in open air by focusing a linearly polarized pulsed beam with a 1 kHz repetition rate, 160 fs pulse duration, and 1030 nm central wavelength to a 43 µm Gaussian spot (1/e2) diameter. The samples were mounted on a motorized, computer-controlled, X-Y-Z translation stage, which enabled scanning (3 mm/s) and positioning out-of-focus to widen the scanning line. For higher precision, the irradiations were conducted at identical conditions, covering a surface of 5 × 5 mm at a pulse power of 2.5 mW (fluence of 0.662 J/cm2) and scan velocity of 3 mm/s with a constant distance between the lines of 15 µm.

In order to determine the effects of chemical composition and morphology on wetting, the drops of distilled water were placed onto both the unmodified and laser-processed surface areas in open air. The volume of each droplet was of 3 µL, controlled by a motorized syringe, while the shape was determined by micro-photographing in the horizontal direction by using Data Physics OCA (optical contact angle) Series device. The values of contact angles were estimated from the images by the utilization of the Gwyddion software [24].

## 3. Results and Discussion

The wetting properties of the surfaces were influenced both by their chemical composition and by their morphology—i.e., corrugation [25,26]. The influence of the different chemical composition is represented by different materials of the topmost layers, while the influence of the morphology is caused by the interaction with the laser. The role of the more complex multilayer structure is represented by using 8×(Zr/Cr/Ti) samples. Exposing the samples to the fs laser beam generated LIPSS on the surfaces, shown in Figure 1. LIPSS on the surface of 15×(Ti/Zr) on Si are shown for two magnifications in Figure 1a,b. LIPSS on the 8×(Zr/Cr/Ti) are presented in Figure 1c,d, and also for two different magnifications. In addition to wavy-patterned LIPSS, the occurrence of nanoparticles with dimensions less than the periods of LIPSS, is also noticeable.

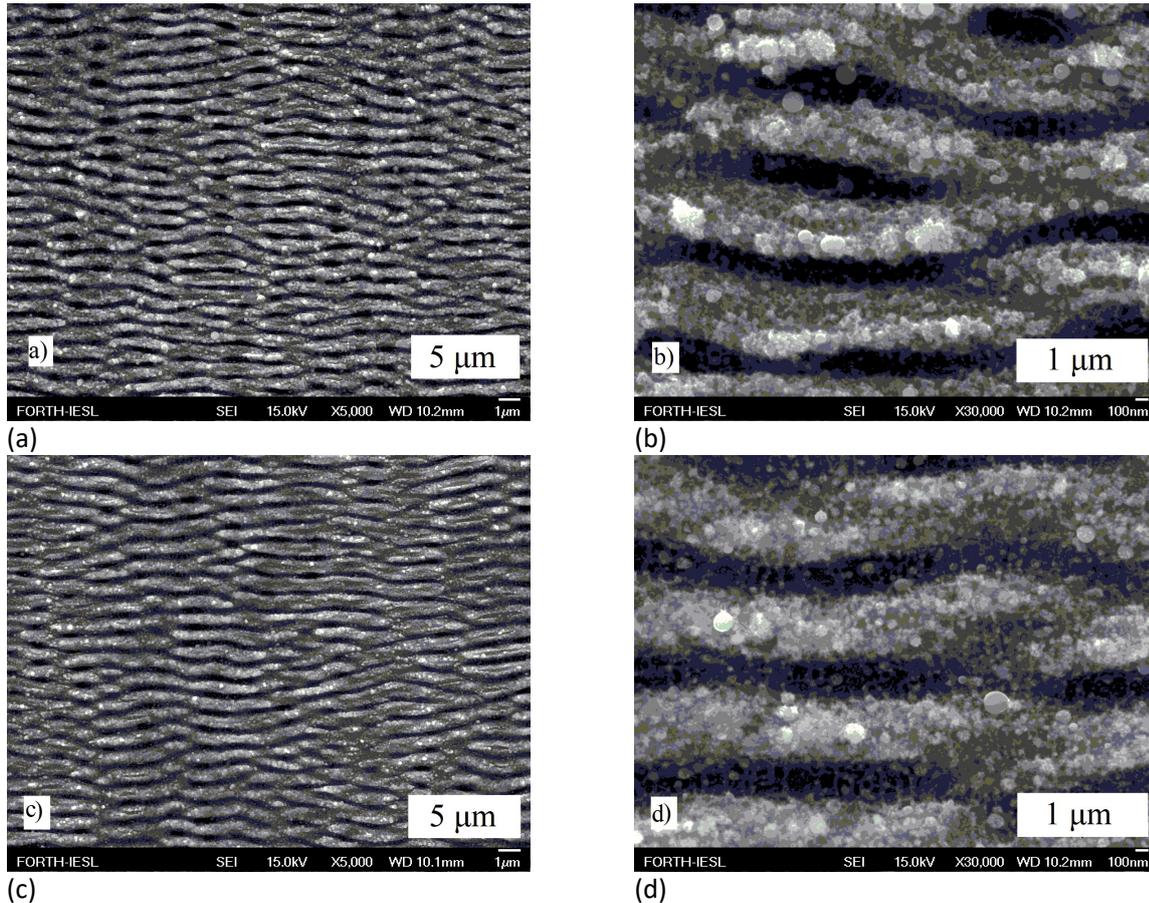

**Figure 1**. SEM micrographs of LIPSS on the surface of the: (a) 15×(Ti/Zr) on Si, magnification 5000×; (b) 15×(Ti/Zr) on Si, magnification 30,000×; (c) 8×(Zr/Cr/Ti) on Si, magnification 5000×; (d) 8×(Zr/Cr/Ti) on Si, magnification 30,000×.

Since the ablation threshold for the multilayer components is lower than the applied fluence, the ablation of the multilayer samples was expected [27]. The experiment with similar conditions (beam, environment, material), created similar ablative LIPSS, while the remaining material in the laser-affected area retained the layered structure (alternating Ti and Zr layers) [28]. Due to the similarity in both conditions and of the outcomes (LIPSS) between the two experiments, it is possible to assume that in the presented experiment the material also retained layered structure in laser affected zones and there were no drastic changes in Ti, Zr, Si and O components among the three zones—the center of, the periphery of, and away from the laser-affected zone.

SEM results (Figure 1) reveal that there is a periodicity in the pattern, but the 2DFFT processing of LIPSS micrograph from Figure 1a presented in Figure 2 provides better insight. The appearance of the maxima (inset in Figure 2) shows that the structures are periodic. A clear distinction in the maxima indicates that LIPSS are highly regular, due to the multi-layer structure. By the convenient combination of the materials, the existence of the underneath layer produces a steep change in the layers' temperatures and enables the transfer of thermal energy through the interface and away from the interaction zone [11].

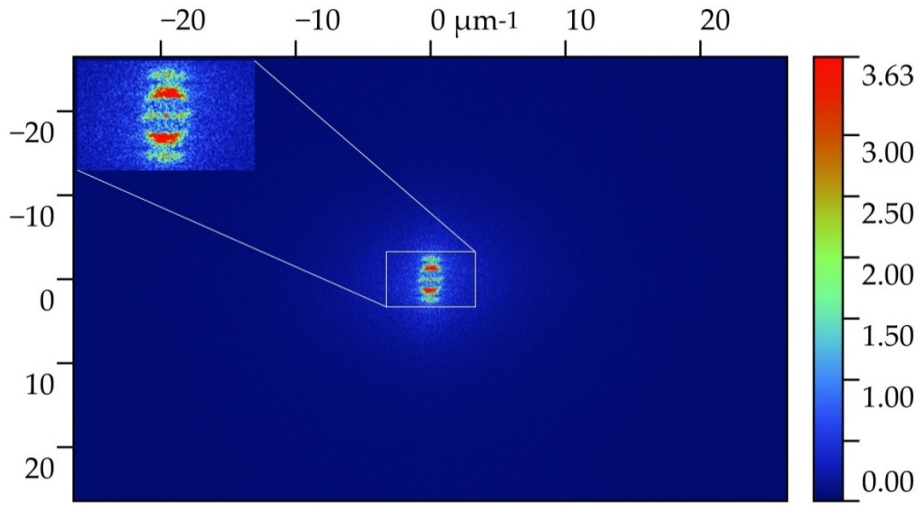

**Figure 2**. Representative 2D-FT of the image shown in Figure 1a. Inset: detailed view of the central part. Intensity scale (bar on the right) is in arbitrary units.

The interaction of fs beam with multilayer thin film materials induces the intermixing of initially separated layers, which could be attributed to the alloying between the components [29]. Due to no drastic changes in the components, not much of the oxides were produced, suggesting that their contribution to the CA value was not significant.

The contact angle of a surface processed by laser beam evolves after the processing. It was shown for some alloys that CA increases to the equilibrium level with a higher rate for lower implemented laser fluence: for alloys containing Ti and fluence of 0.78 J/cm2, it takes a couple of days to be close enough to the equilibrium [16]. The images of water droplets, placed on the surfaces both unexposed and exposed to fs beams, are presented in Figure 3. For 15×(Ti/Zr), the droplets are shown in Figure 3a (on unexposed surface) and Figure 3b (on exposed surface). Different shapes indicate the influence of the laser-induced change of the surface, which is twofold: both in the morphology and in the chemical composition. Similar holds for the droplets on the 15×(Zr/Ti) unexposed (Figure 3c) and exposed (Figure 3d) surfaces, as well as on the 8×(Zr/Cr/Ti) unexposed (Figure 3e) and exposed (Figure 3f) surfaces. On exposed surfaces, the droplets were placed 180 days after the exposition (15×(Ti/Zr)) and 2 days after the exposition (15×(Zr/Ti) and 8×(Zr/Cr/Ti)).

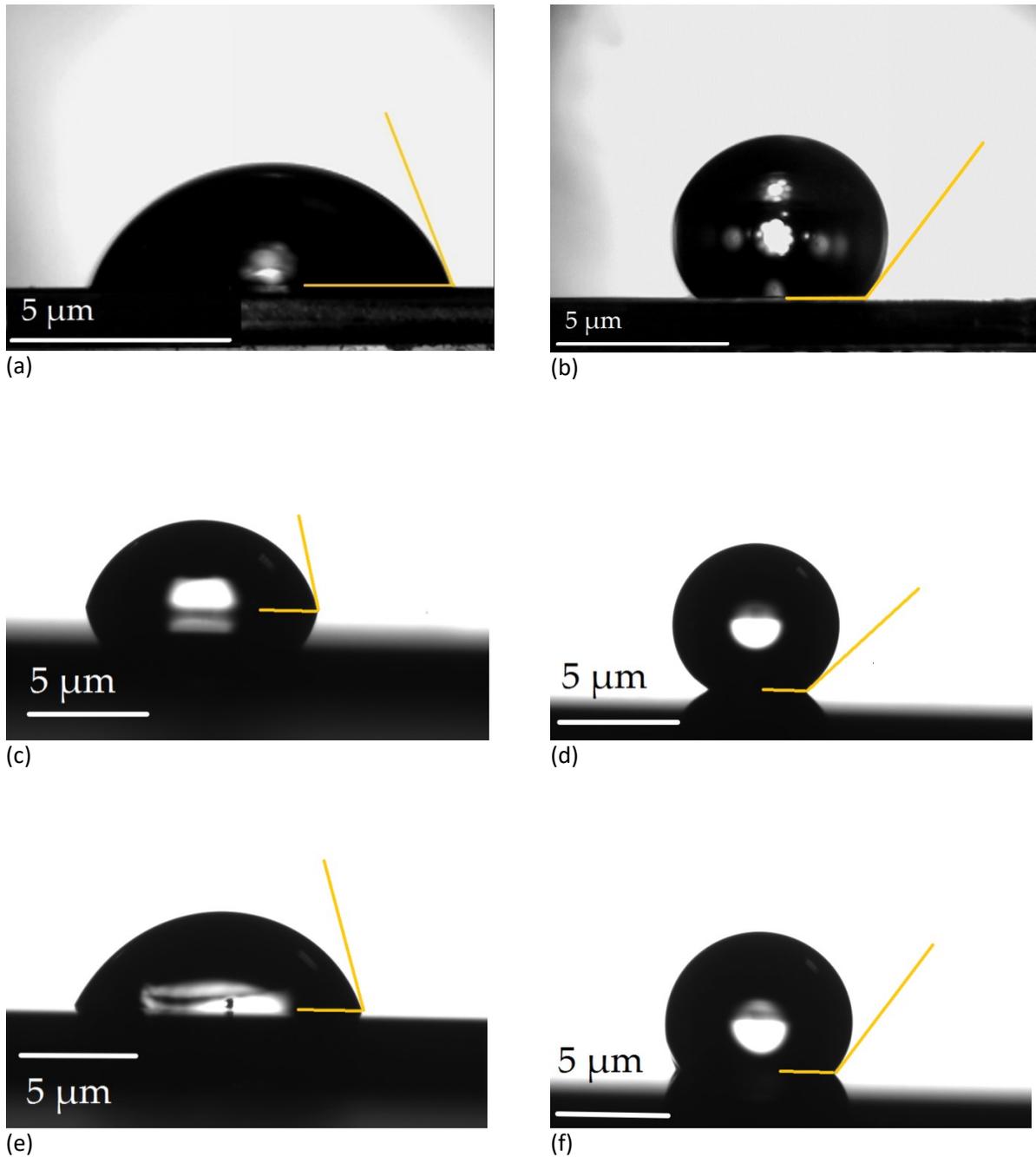

**Figure 3**. The image of the droplet of distilled water on the surface of: (a) 15×(Ti/Zr) on Si (Ti is the topmost layer), as-deposited, the CA = 72.11°; (b) 15×(Ti/Zr) on Si (Ti is the topmost layer), laser-modified, the CA = 137.15°; (c) 15×(Zr/Ti) on Si (Zr is the topmost layer), as-deposited, the CA = 77.29°; (d) 15×(Zr/Ti) on Si (Zr is the topmost layer), laser-modified, the CA = 144.49°; (e) 8×(Zr/Cr/Ti) on Si (Zr is the topmost layer), as-deposited, the CA = 68.10°; (f) 8×(Zr/Cr/Ti) on Si (Zr is the topmost layer), laser-modified, the CA = 123.45°. White bar denotes 5 μm. Golden lines are added for the visualization.

As a consequence of ultra-short laser beam interaction with metal surfaces in open air, the surface plasmon polariton (SPP) occurred at the interface between the metal and the dielectric (air) surfaces, inducing periodic distribution and accumulation of the energy on the surface. Because fs pulse duration is shorter than the characteristic relaxation time of the material, energy deposition is shock-like and leads to the ablation of the material, leaving parallel trenches—LIPSS. The spatial period of LIPSS is less than the incoming wavelength. The heat load to the surrounding material was minimal. The ablated material was dispersed away in the form of nanoparticles with the dimensions of up to 100 nm. In this way, the deposited energy was mostly invested in the ablation and not in heating, which implies that, for fs interaction, the creation of a thick oxide layer did not take place.

The values of the LIPSS periods estimated from the micrographs, as well as measured values of contact angles of both surfaces, are summarized in Table 1. The period matches between 72 and 77% of the value of the implemented wavelength. The uniformly distributed LIPSS are oriented in the normal to the polarization directions of the incoming beam and their period is not close to the used laser wavelength for all observed samples. These types of LIPSS originate from the interference of the incident laser beam with surface electromagnetic wave excited during the laser treatment [8]. The CA values indicate that non-corrugated surfaces are hydrophilic. Similar CA values for surfaces unmodified by the laser suggest that they are not much influenced by the difference in the composition. The difference in CA between the untreated samples of Zr/Ti and Zr/Cr/Ti could be explained by long-range wetting transparency [30]. For surfaces modified by laser beam, the values also do not differ much. However, a strong increase after the interaction by the fs laser beam suggests it is dominantly caused by the structural change—the occurrence of LIPSS—and not by the composition change. Different to the case where a 40° rise in hydrophobicity of Ti surface was obtained by μm corrugations [17], better results in this work were obtained by nm corrugations. In open air ambience, all transition metals (except a few) are covered with a thin native oxide layer [31]. Different values of CA in the literature are the consequence of different preparation methods, which lead to differences in the chemical composition on the interface. The Thin zirconium oxide layer (80–90 nm) on the SiO2 substrate shows hydrophilicity (small values of CA (hydrophilicity), but thicker layers show higher values of CA [32,33]. Oxides of titanium have various values of water contact angles [34]. Due to long-range wetting transparency, the water contact angle of the MgF2 layer, thinner than 100 nm, becomes highly insensitive of its chemical composition and its value approaches the value of the CA of the underlying materials: glass, Au and Au/MgF2 metamaterial [30]. Therefore, it could be assumed that very thin (<100 nm) layers of metal-oxides did not contribute to the water CA value. Due to the similarity in the experiments [28], the change in the components was small and due to the forming of LIPSS by fs ablation (energy invested in ablation, not in heating), not many of the oxides were produced and their contribution to the CA value was not significant. The CA changed mostly due to the corrugations and not the presence of oxides.

**Table 1**. The LIPSS period of the sample materials and the CA values.

| Material | LIPSS Period (nm) | CA (°) | | ΔCA (°) | CA Increase (%) |
|---|---|---|---|---|---|
| | | untreated surface | corrugated surface | | |
| 15×(Ti/Zr) | 740 | 72.11 ± 3.32 | 137.15 ± 11.63 | 65.04 | 90 |
| 15×(Zr/Ti) | 740 | 77.29 ± 1.71 | 144.49 ± 14.97 | 67.20 | 87 |
| 8×(Zr/Cr/Ti) | 790 | 68.10 ± 6.41 | 123.45 ± 25.26 | 55.35 | 81 |

In this view, all the samples should have a similar increase in the CA after the interaction. However, there is a small difference which would mean that slightly different amounts of Zr and Ti components are removed by laser treatment depending on the initial order of the layers. Here, a significant mixture of the materials did not take place and spatial distribution retained a layered structure, which resulted in a similar increase in the CA angle.

From the SEM images (Figure 1), the Gwyddion program was used to graphically extract the profiles along the selected lines perpendicular to the direction of the LIPSS, and from them to estimate the filling factors—the ratio between the ripple width and the spatial period. For 15×(Ti/Zr), it is ~0.6, while for 8×(Zr/Cr/Ti), it is ~0.72. It is possible to assume that greater filling factor leads to greater Casimir pressure for the same distance between the surfaces, while the corrugation height does not play a significant role [23]. The topmost surface is covered with a very thin oxide layer, because the ambience is open air. Laser interaction will enforce the oxidation, but only to the saturation level [35]. However, the results presented in Table 1 show that whichever film structure we have at the beginning, after the interaction with the laser, the creation of the nanoscale corrugation causes an increase in CA. The higher filling factor for the samples with Zr as the topmost layer should cause the CA values' relatively more significant change than for those with Ti on top. This was observed for 15×(Zr/Ti), but in the case of 15×(Zr/Cr/Ti), the long-range wetting transparency [30] means that CA is also affected by the presence of a thin Cr layer.

While LIPSS are anisotropic, the isotropy of the contact angles was not investigated. The direction of LIPSS (sample orientation) relative to the direction of recording (camera orientation) was arbitrary for each sample and undetermined. No significant difference in the left and right contact angles can be seen in Tas Figure 3b,d,f, meaning either LIPSS and recording directions were set parallel by chance for each sample or the droplets were not significantly anisotropic. To match the directions of LIPSS and camera recording to be parallel by chance is less likely, indicating that the droplets are not significantly anisotropic. Though the anisotropy of the surface structures influences the anisotropy of the CA, with the smaller width of the structures (ripples itself), the influence is smaller [36]. Very small widths of the ripples (~700 nm) in the presented experiment suggest that the influence of the structure anisotropy (LIPSS) to the droplet anisotropy is not significant.

Corrugation is responsible for hydrophobicity (wetting). Classical theory demands the size of corrugation to be above micrometer level in order to achieve super-hydrophobicity. If we assume that a multilayer structure can be imagined as an optical cavity, quantum effects cause the rise of super-hydrophobicity [21,22]. Recently, experimental proof of the impact of nano-corrugation to wetting properties exploiting the Casimir effect has been published [37]. The effect of small corrugations and quantum effects to superhydrophobicity is phenomenologically confirmed. In our work, it has been shown that, for laser generated nano-corrugation, observed wetting properties can be explained only if quantum effects are taken into account. Corrugations of nanoscopic level dramatically affect the Van der Waals interaction energy (through quantum vacuum photon modes) and thus the wetting contact angle is modified. Up to now, to achieve super-hydrophobicity, different coatings and dozens of micro-size structures were described elsewhere [38,39]. For thin film materials, the wetting angle is nearly insensitive to the chemical nature of the immediate substrate and, by the forming of hyperbolic (plasmonic) metamaterials, the effective refractive index changed as well as the contact angle [30].

Here, the authors are the first, to our knowledge, to report significant change in the wetting properties by subwavelength corrugation. The increase in hydrophobicity caused by fs nano-patterning experimentally confirms the model developed by Dellieu et al. [21,22].

Furthermore, the presented study goes beyond the proposed model, since super-hydrophobicity is achieved on the metallic surface (not only the molecular solid). In the end, the possibility of attaining super-hydrophobicity using submicron corrugation by controlling quantum vacuum modes opens large almost unlimited fundamental and technological applications. While the responsibility for the change in the dispersion component is doubtful, the corrugation is responsible for all the changes in the CA. However, the magnitude of the effects is too big for such small corrugations, which can be explained with the abovementioned non-classical approach [21,22]. Untreated metal surfaces are generally hydrophilic, but in this experiment they changed to highly hydrophobic by the implementation of sub-micrometer-sized corrugation.

## 4. Conclusions

We have investigated the wetting of laser modified multilayer thin films of different metal materials by using contact angle measurements. The values do not differ much for different materials (both as-deposited and laser modified) but significantly change after the laser interaction and the formation of LIPSS. Here, due to nanometer-sized structuring, wetting cannot be explained without taking quantum effects into consideration. To our actual knowledge, this is the first example of significant wetting change by nanostructures and it is in good agreement with previous publication [21,22], which explains wetting by quantum vacuum modes. This suggests that the increase in the hydrophobicity is related to the morphological changes in the cavity. The presented results extend the application of the model previously developed [21,22] for molecular solids, and show that the nanocorrugation of metal surfaces can be used to significantly affect the wetting. While the influence of quantum modes freezing to superhydrophobicity has been phenomenologically confirmed [37], we have shown that observed wetting properties of surfaces with fs laser generated non-corrugations can be explained only if quantum effects are taken into account. The potential applications of the presented research could be numerous.

Author Contributions: The contributions are as follows: conceptualization, B.K. and A.G.K.; methodology, B.K. and S.P.; validation, B.K. and S.P.; investigation, A.M. and S.P.; resources, E.S.; data curation, A.M. and S.P.; writing—original draft preparation, A.G.K.; writing—review and editing, A.G.K., S.P., A.M., E.S., D.P. and B.K.; visualization, A.G.K. and S.P.; supervision, E.S.; project administration, D.P., E.S. All authors have read and agreed to the published version of the manuscript.

Funding: This research was funded by Ministry of Education, Science and Technological Development of the Republic of Serbia, grant numbers III45016 and OI171038 and by EU-H2020 European Research Infrastructures NFFA-Europe research and innovation program under grant agreement No. 654360.

Acknowledgments: The authors are also grateful to Danilo Kisić and dr Davor Peruško, both of the Institute of Nuclear Sciences "Vinča" (University of Belgrade) for preparing the multilayer thin metal film samples. B.K. acknowledges support of F.R.S.-FNRS.

Conflicts of Interest: The authors declare no conflict of interest. The funders had no role in the design of the study; in the collection, analyses, or interpretation of data; in the writing of the manuscript, or in the decision to publish the results.